\documentstyle[12pt,epsfig]{article}
\title{LHC(CMS) SUSY discovery potential for nonuniversal gaugino and squark 
masses and the determination of the effective SUSY scale}
\author{\large S.I.~Bityukov~$^1$, N.V.~Krasnikov  \\[3mm]
\em Institute for Nuclear Research RAS, \\
\em Moscow, 117312, Russia  }
\date{}
\begin{document}
\maketitle
\begin{abstract}
We review the results on the investigation of the LHC(CMS) SUSY discovery 
potential for the case of nonuniversal gaugino and squark masses.  

\end{abstract}
\vspace{1cm}
\bigskip

\noindent
\rule{3cm}{0.5pt}\\
$^1$~~Institute for High Energy Physics, Protvino, Russia

\bigskip

One of the LHC supergoals is the discovery of the supersymmetry.
In particular, it is very important to investigate a possibility 
to discover strongly interacting superparticles (squarks and gluino). 
In Ref.\cite{1} (see also Refs.\cite{2}) the LHC squark and gluino 
discovery potential has been investigated within the minimal 
SUGRA-MSSM   
 framework \cite{3} where all 
sparticle masses are determined mainly by two parameters: $m_0$ (common
squark and slepton mass at GUT scale) and $m_{1 \over 2}$ (common
gaugino mass at GUT scale). 
The signature used for the search for squarks and gluino  at LHC is 
$(n \geq 0)~isolated~leptons + (m \geq 2) jets + E^{miss}_T$ events. 
The conclusion of Ref.~\cite{1}  is that LHC is able 
to detect squarks and gluino with masses up to $(2 - 2.5)~TeV$.

Despite the simplicity of the  SUGRA-MSSM framework it is a very particular
model. The mass formulae for sparticles in  SUGRA-MSSM model are derived
under the assumption that at GUT scale ($M_{GUT} \approx 2 \cdot 10^{16}$~GeV) 
soft supersymmetry breaking terms are universal. However, in general,
we can expect that real sparticle masses can differ in a drastic way 
from sparticle masses pattern of SUGRA-MSSM model due to many reasons,
see for instance refs.~\cite{4,5,6,7}.
Therefore, it is more appropriate to investigate the LHC SUSY discovery 
potential in a model-independent way for general case of nonuniversal 
gaugino and squark masses.

In this report we review our previous results \cite{8}-\cite{11} on 
the investigation  of the LHC(CMS) SUSY discovery potential for 
the case of nonuniversal squark and gaugino masses. We also give 
some new results on the determination of the effective SUSY scale 
for models with nonuniversal gaugino masses.

The decays of squarks and gluinos depend on the relation among 
squark and gluino masses. For $m_{\tilde{q}} > m_{\tilde{g}}$ squarks 
decay mainly into gluino and quarks  and  gluino decays mainly 
into a quark-antiquark pair and a gaugino. For    
$m_{\tilde{q}} < m_{\tilde{g}}$ gluino decays mainly into squarks and quarks, 
whereas squarks decay mainly into quarks and a gaugino. The next-to-lightest 
gauginos have several leptonic decay modes giving a lepton 
and missing energy, for instance:
\begin{equation}
\tilde{\chi}^{\pm}_1 \rightarrow \tilde{\chi}^{0}_1 + l^{\pm} + \nu,
\end{equation}
\begin{equation}
\tilde{\chi}^{0}_2 \rightarrow \tilde{\chi}^{0}_1 + l^{\pm} l^{\mp}
\end{equation}
As a result of chargino and second neutralino leptonic decays,
besides the classical signature $(n \geq 2,3,4)~jets~plus~E_T^{miss}$ 
signatures such as    $(n \geq 2,3,4)~jets~plus~(m \geq 1~ 
leptons) ~plus~E_T^{miss}$ with leptons and jets in final state arise.

In our concrete calculations all SUSY processes, with full particle 
spectrum, couplings, production cross section and decays have been generated 
with ISASUSY code \cite{12}. The Standard Model backgrounds have been 
generated by PYTHIA code \cite{13}. Our simulations have been made with 
parametrized detector responses based on detailed detector simulations. 
Namely, the CMS(Compact Muon Solenoid) detector fast simulation code 
CMSJET \cite{14} has been used. It appears that the following 
SM processes give the main contribution to the background:

$W ~+~jets,~Z~+~jets,~t\bar{t},~WZ, ~ZZ, ~b\bar{b}~and
~QCD(2 \rightarrow 2)~reactions$ .

As it has been mentioned previously in our study  
we have considered as signatures $(n \geq l)$ jets plus $(m \geq k)$ 
isolated leptons plus $E^{miss}_T$, where $l = 2,3,4$ and 
$k = 0,1,2,3,4$. More detailed information on the used cuts can be found 
in Refs.\cite{8}-\cite{11}.      
 
For squark and gluino pair production we have considered 4 different 
kinematic regions:

\begin{itemize}

\item 

A. $m_{\tilde{g}} \gg m_{\tilde{q}}$

\item

B. $m_{\tilde{q}} \gg m_{\tilde{g}}$

\item

C. $m_{\tilde{q}} \sim m_{\tilde{g}}$, $m_{\tilde{q}} >  m_{\tilde{g}}$

\item

D. First 2 squark generations are heavy $m_{\tilde{q}_{1,2}} \gg 1~TeV$. 
Only 3 rd squark generation is relatively light.

\end{itemize}

We have found that the LHC(CMS) SUSY discovery potential depends 
rather strongly on the relation among $m_{\tilde{\chi}^0_1}$ 
and $min(m_{\tilde{g}}, m_{\tilde{q}})$ and it decreases with the 
increase of the LSP mass    $m_{\tilde{\chi}^0_1}$. For LSP 
mass $m_{\tilde{\chi}^0_1}$ close to  $min(m_{\tilde{g}}, m_{\tilde{q}})$ 
it is possible to detect SUSY for 
 $min(m_{\tilde{g}}, m_{\tilde{q}}) \leq (1.2 - 1.5)~TeV$ (cases A,B,C).
For the case D it is possible to detect SUSY with $m_{\tilde{q}_{1,2}}
\leq 800~GeV$. Our results are presented in Figs.(1-5).

Gluinos and squarks are strongly produced at the LHC, so it is important 
to find a variable that measures the produced mass for events with 
missing transverse energy. In Ref.\cite{15}
\footnote{See also recent  Ref.\cite{16} on this subject.}
 a variable that works rather 
well within SUGRA-MSSM model has been defined as the sum of the missing 
energy and the $p_T$'s of the first four jets,
\begin{equation}
M_{eff} = E_T^{miss} + p_{T,1} + p_{T,2} + p_{T,3} + p_{T,4}
\end{equation}
As it has been mentioned  in Ref.\cite{15} in SUGRA-MSSM model 
the distribution of SUSY cross section $\frac{d\sigma_{SUSY}}{dM_{eff}}$ 
has maximum at some point $M_{SUSY} \equiv M_{peak}$
and the value of 
$M_{SUSY}$ coincides within 10 percent accuracy with 
$min(m_{\tilde{g}}, m_{\tilde{q}})$, namely  
\begin{equation}
M_{SUSY} \equiv M_{peak} \approx  min(m_{\tilde{g}}, m_{\tilde{q}}) 
\end{equation}

We have investigated the dependence of $M_{SUSY}$ on squark, gluino and 
LSP masses in general case of arbitrary relations among LSP, squark and 
gluino masses. Our results are presented in Figs.(6-8). As it follows from our 
results in general case for LSP mass $m_{\tilde{\chi}^0_1} 
\leq    m_{\tilde{g}}$ there is no linear dependence of $M_{SUSY}$ on 
gluino or squark masses. The value of  $M_{SUSY}$ is nonlinear and nontrivial 
function on gluino, squark and LSP masses. So, in general case it would be 
rather nontrivial to determine more or less accurately the squark and 
gluino masses with the help of  $M_{SUSY}$.

In conclusion we would like to stress that even for the most difficult 
case when LSP mass is rather heavy and close to squark or gluino masses 
the LHC is able to discover SUSY for the most interesting from 
the theoretical point of view case when sparticle masses are lighter than 
$1~TeV$.

This work has been supported by CERN-INTAS Grant 99-0377 and 
RFFI Grant N02-01-00601.


\begin{figure}[H]

\centerline{\epsfig{file=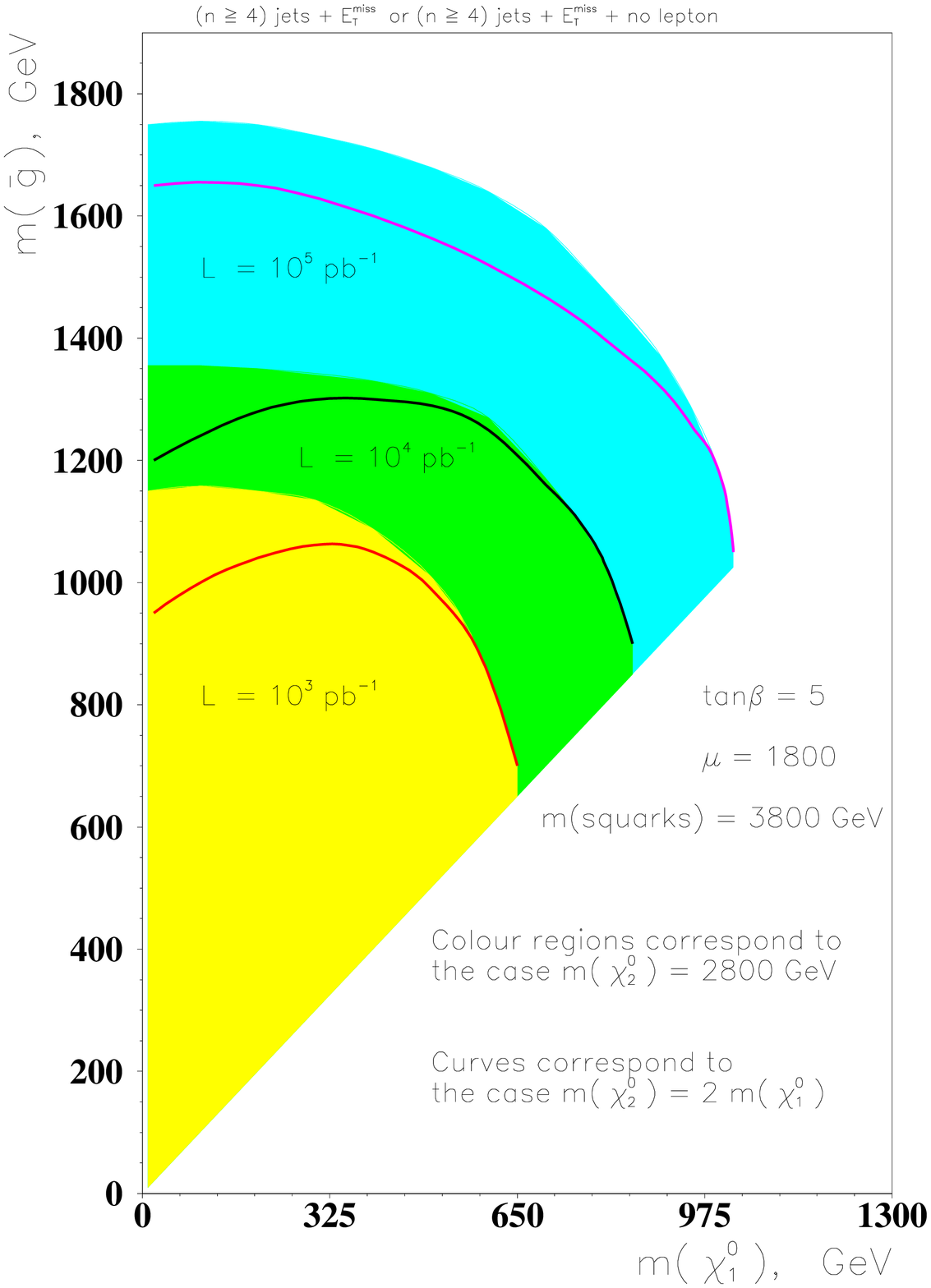,width=14cm}}

\vspace*{-0.4cm}

\caption{\small CMS discovery potential  for different values
of $m_{\tilde \chi_1^0}$ and $m_{\tilde g}$ in the case of 
$m_{\tilde g} > m_{\tilde \chi^0_1}$.}
\label{fig.1}

\end{figure}

\begin{figure}[H]

\centerline{\epsfig{file=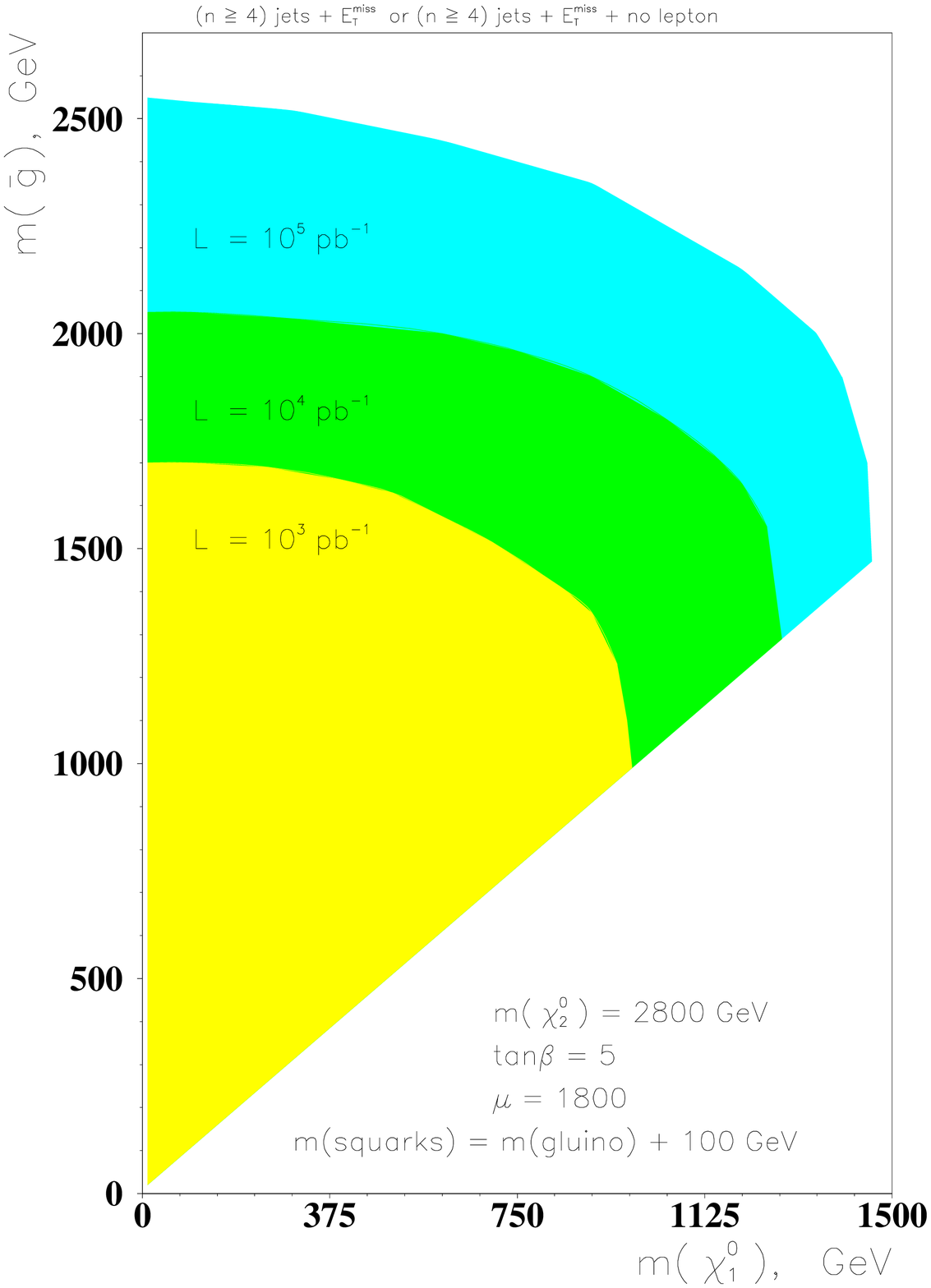,width=14cm}}

\vspace*{-0.4cm}

\caption{\small CMS discovery potential  for different values
of $m_{\tilde \chi_1^0}$ and $m_{\tilde g}$ in the case of 
$m_{\tilde g} > m_{\tilde \chi^0_1}$}
\label{fig.2}

\end{figure}

\begin{figure}[H]

\centerline{\epsfig{file=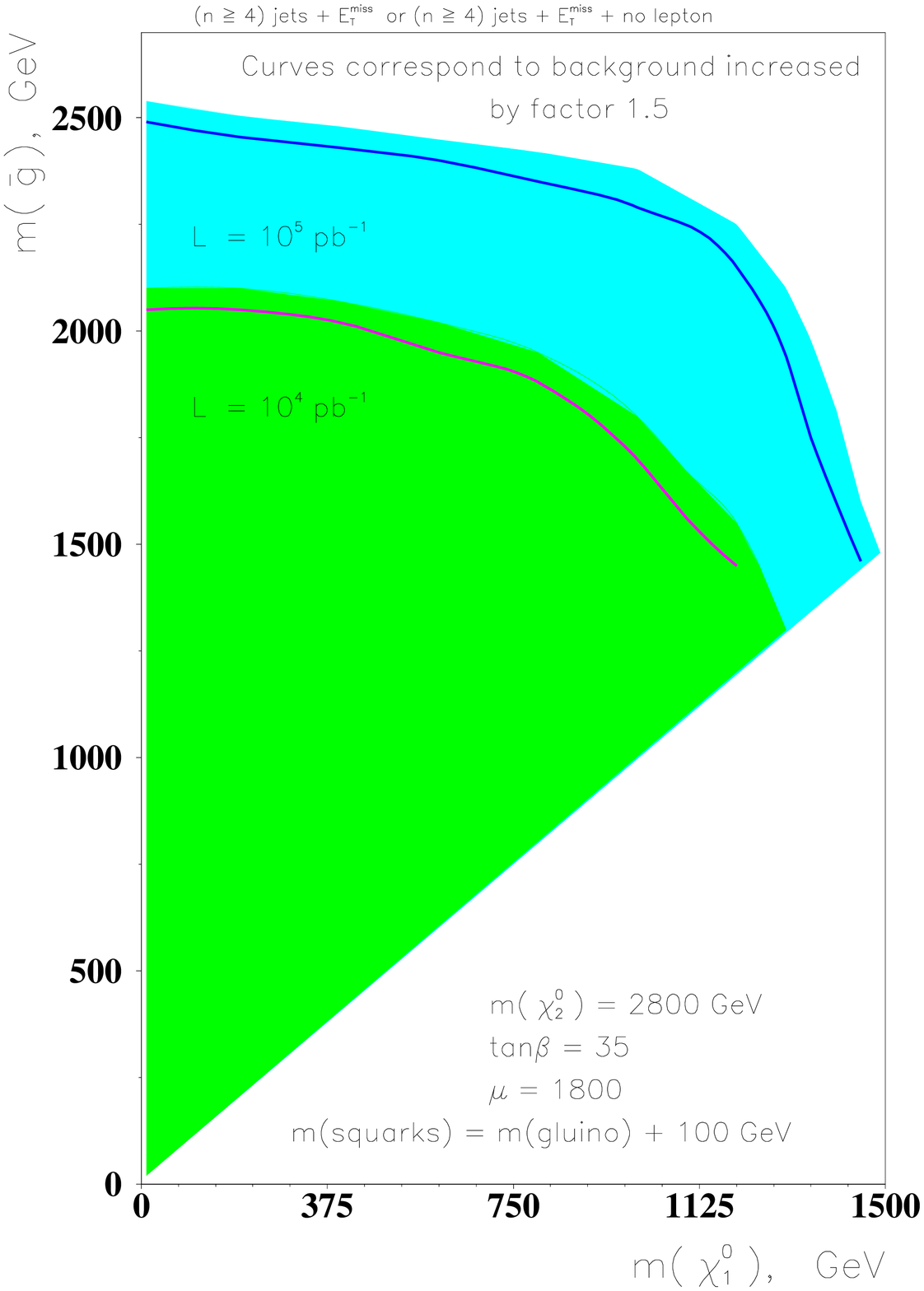,width=14cm}}

\vspace*{-0.4cm}

\caption{\small CMS discovery potential  for different values
of $m_{\tilde \chi_1^0}$ and $m_{\tilde g}$ in the case of 
$m_{\tilde g} > m_{\tilde \chi^0_1}$.}
\label{fig.3}

\end{figure}

\begin{figure}[H]

\centerline{\epsfig{file=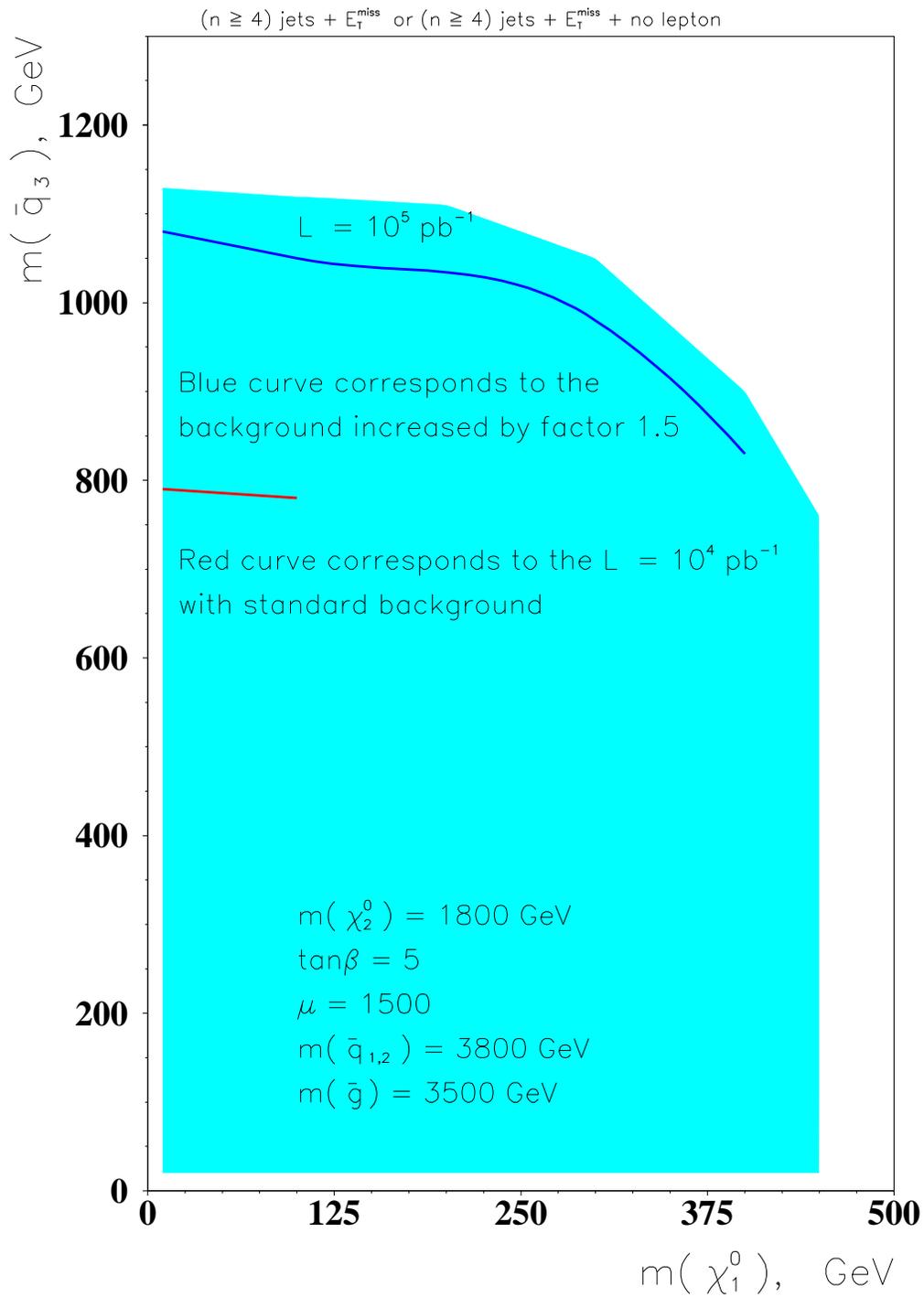,width=14cm}}

\vspace*{-0.4cm}

\caption{\small CMS discovery potential  for different values
of $m_{\tilde \chi_1^0}$ and $m_{\tilde q_3}$ in the case of 
$m_{\tilde q_{1,2}} \gg m_{\tilde q_3}$.}
\label{fig.4}

\end{figure}

\begin{figure}[H]

\centerline{\epsfig{file=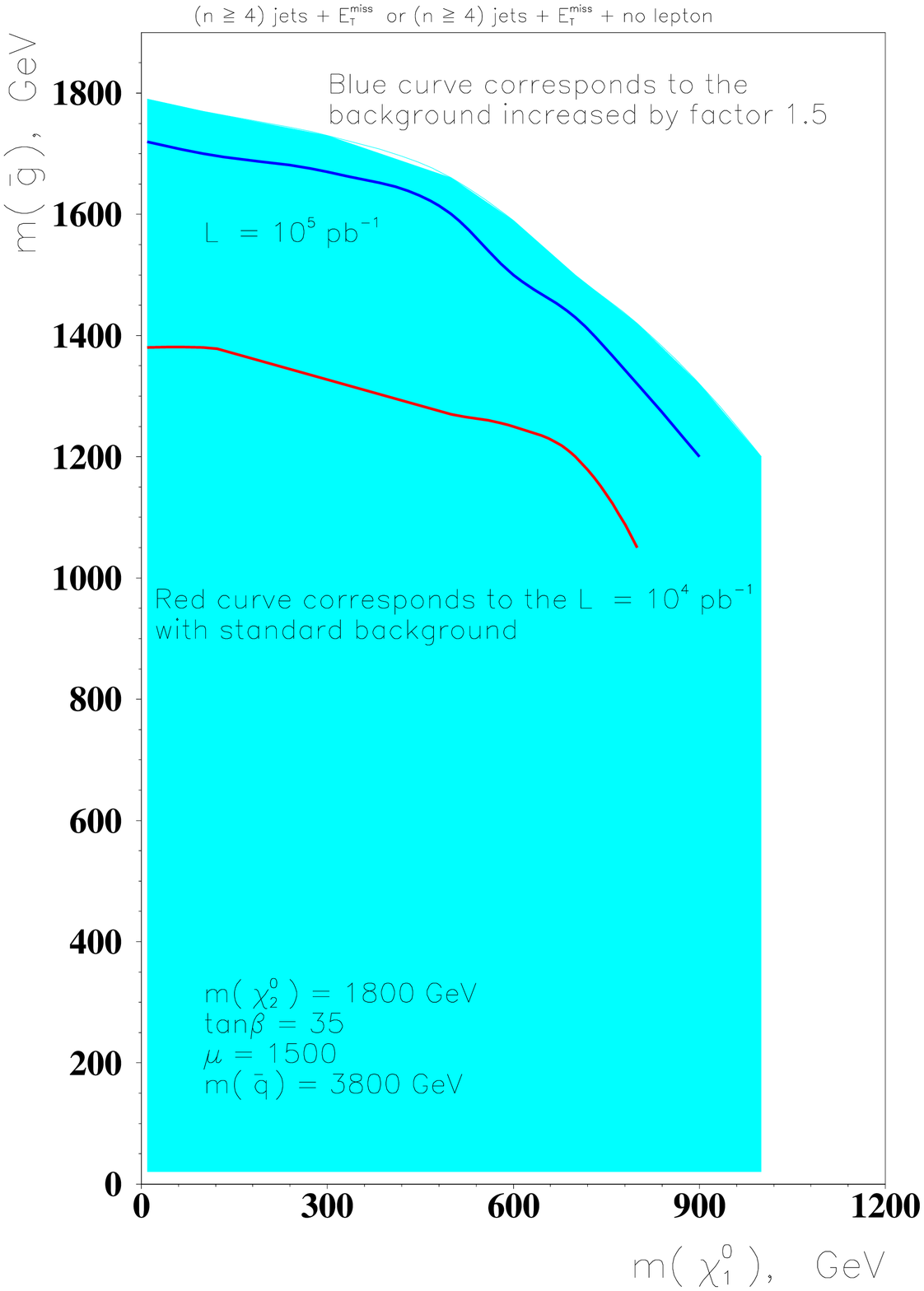,width=14cm}}

\vspace*{-0.4cm}

\caption{\small CMS discovery potential  for different values
of $m_{\tilde \chi_1^0}$ and $m_{\tilde g}$ in the case of 
$m_{\tilde q} \gg m_{\tilde g}$.}
\label{fig.5}

\end{figure}

\begin{figure}[H]

\centerline{\epsfig{file=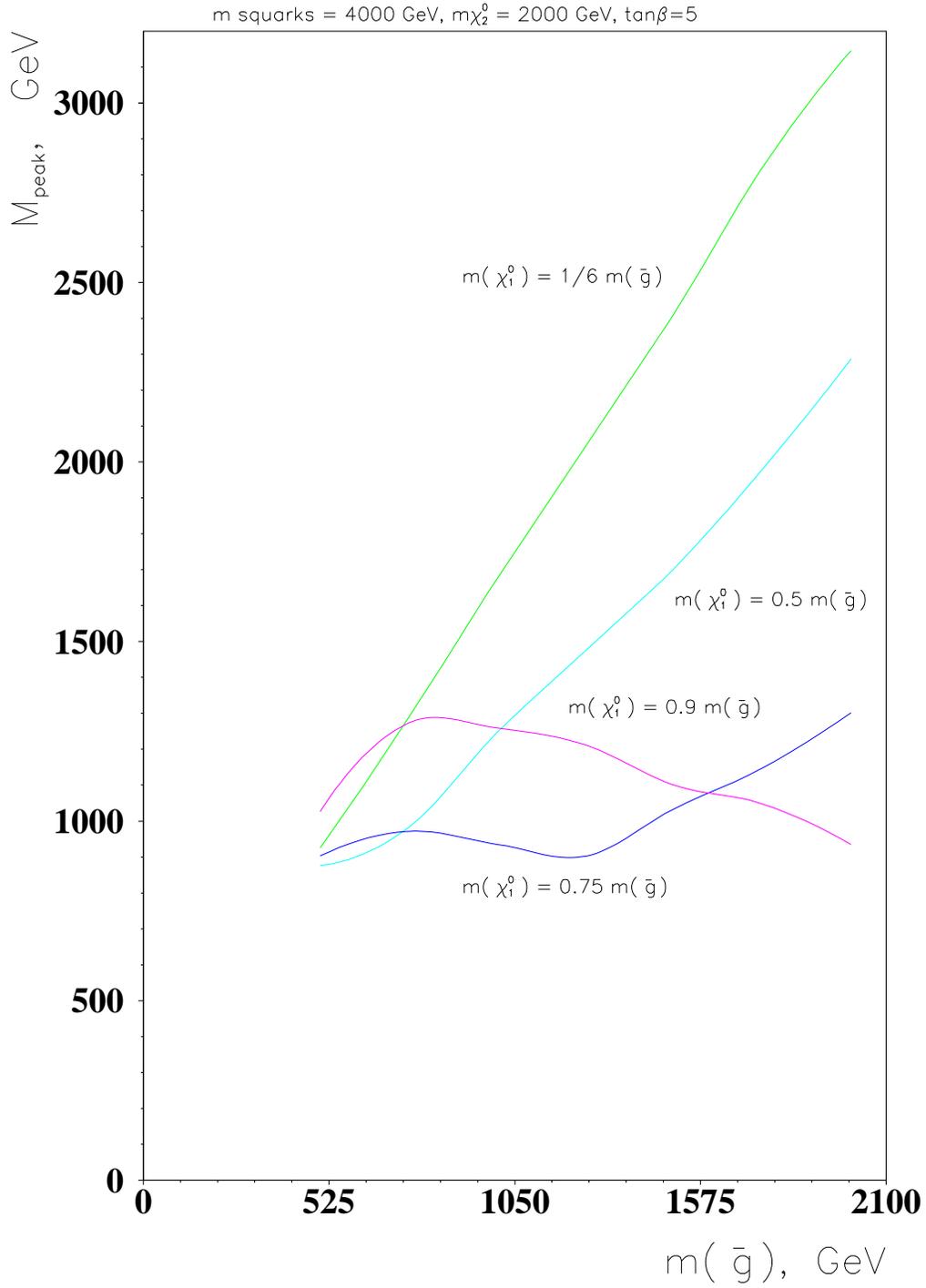,width=14cm}}
\vspace*{-0.4cm}

\caption{\small The dependence of $M_{susy} \equiv M_{peak}$ on the 
gluino mass for different values of $m(\chi^0_1)$ .}
\label{fig.6}

\end{figure}

\begin{figure}[H]

\centerline{\epsfig{file=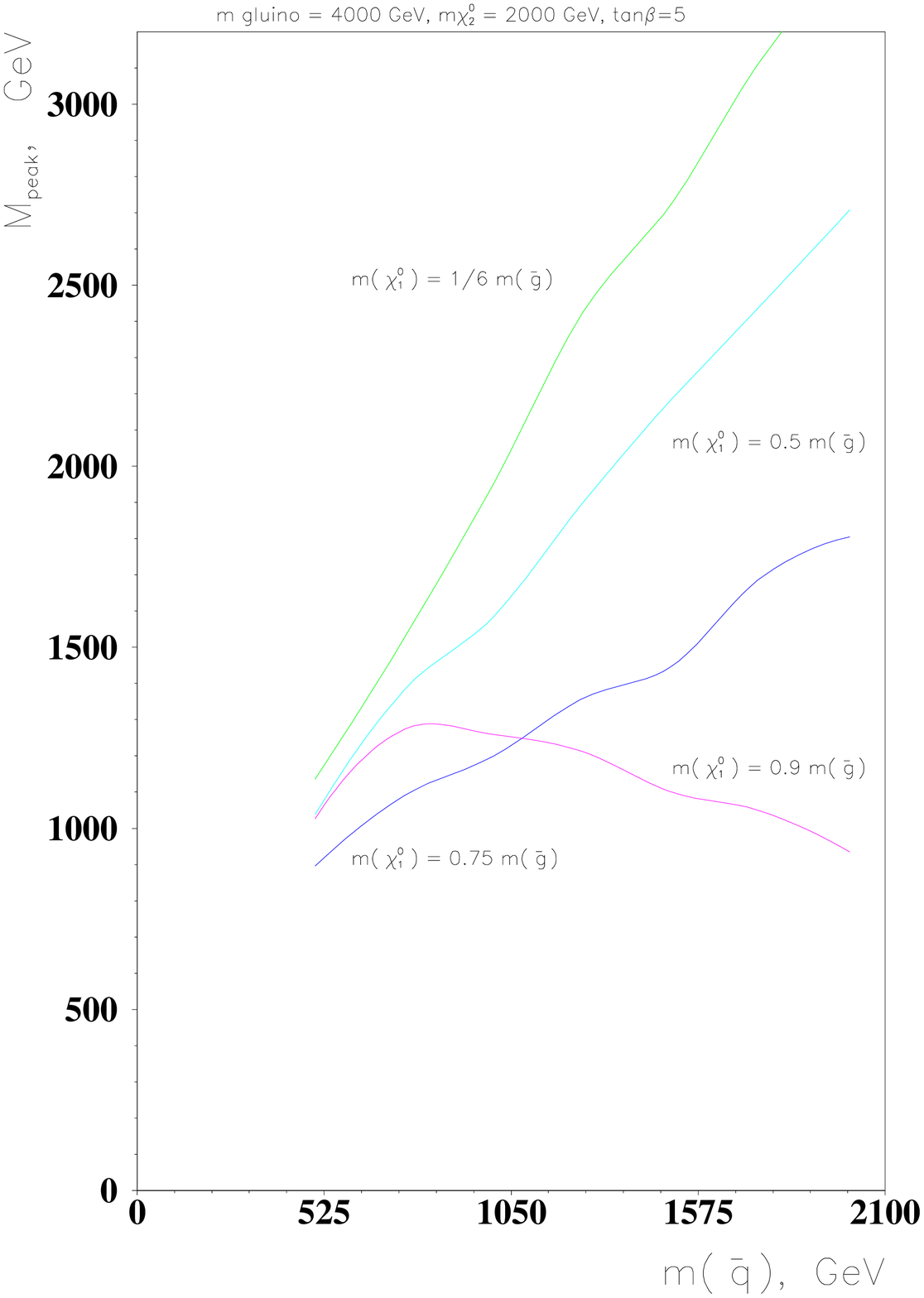,width=14cm}}
\vspace*{-0.4cm}

\caption{\small The dependence of $M_{susy} \equiv M_{peak}$ on the 
squark mass for different values of $m(\chi^0_1)$.}
\label{fig.7}

\end{figure}

\begin{figure}[H]

\centerline{\epsfig{file=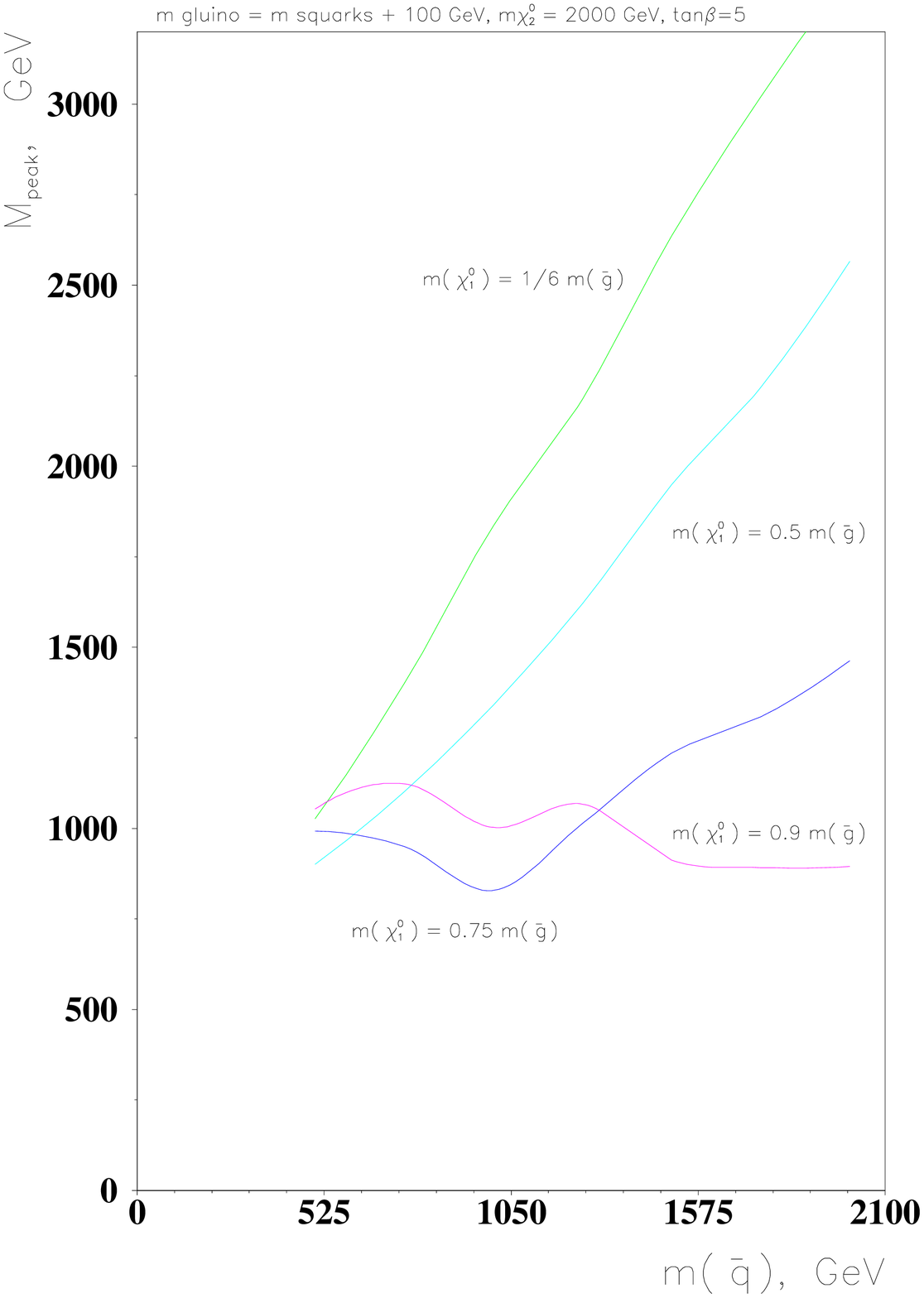,width=14cm}}
\vspace*{-0.4cm}

\caption{\small The dependence of the  $M_{susy} \equiv M_{peak}$ 
on the squark mass for different values of $m(\chi^0_1)$ .}
\label{fig.8}

\end{figure}

\end{document}